\pdfoutput=1

\documentclass[11pt]{article}

\usepackage[]{acl}

\usepackage{times}
\usepackage{latexsym}
\usepackage{amsmath}

\usepackage[T1]{fontenc}

\usepackage[utf8]{inputenc}

\usepackage{microtype}
\usepackage{booktabs}
\usepackage{multirow}
\usepackage{multicol}
\usepackage{graphicx}
\usepackage{makecell}
\usepackage{amsmath}
\usepackage{amssymb}
\usepackage{algorithm}
\usepackage{algpseudocode}
\usepackage{xspace}
\usepackage{pdfpages}
\usepackage{pgfplots}
\usepackage{tikz}
\usepackage{subcaption}

\author{Zehan Li, Yanzhao Zhang, Dingkun Long, Pengjun Xie \\
  Alibaba Group \\
  \texttt{\{lizehan.lzh,zhangyanzhao.zyz\}@alibaba-inc.com} \\
  \texttt{\{dingkun.ldk,pengjun.xpj\}@alibaba-inc.com} \\
}

\newcommand{\cmark}{\checkmark}

\usepackage{tcolorbox}
\usepackage{colortbl,array,xcolor}
\usepackage{soul}
\newcommand{\hlc}[2][yellow]{{%
    \colorlet{foo}{#1}%
    \sethlcolor{foo}\hl{#2}}%
}

\newcommand\ours{\textsc{CDMAE}\xspace}

\title{Instructions for *ACL Proceedings}
\title{Challenging Decoder helps in Masked Auto-Encoder Pre-training \\ for Dense Passage Retrieval}

\begin{document}
\maketitle
\begin{abstract}
Recently, various studies have been directed towards exploring dense passage retrieval techniques employing pre-trained language models, among which the masked auto-encoder (MAE) pre-training architecture has emerged as the most promising. The conventional MAE framework relies on leveraging the passage reconstruction of decoder to bolster the text representation ability of encoder, thereby enhancing the performance of resulting dense retrieval systems. Within the context of building the representation ability of the encoder through passage reconstruction of decoder, it is reasonable to postulate that a ``more demanding'' decoder will necessitate a corresponding increase in the encoder's ability. To this end, we propose a novel token importance aware masking strategy based on pointwise mutual information to intensify the challenge of the decoder. Importantly, our approach can be implemented in an unsupervised manner, without adding additional expenses to the pre-training phase. Our experiments verify that the proposed method is both effective and robust on large-scale supervised passage retrieval datasets and out-of-domain zero-shot retrieval benchmarks.

\end{abstract}

\section{Introduction}
Passage retrieval is a core sub-task in various downstream applications, such as open-domain question answering~\cite{karpukhin-etal-2020-dense,Qu2021RocketQAAO,Zhu2021AdaptiveIS}, conversational systems~\cite{Yu2021FewShotCD} and web search~\cite{Lin2021PretrainedTF,Fan2021PretrainingMI,Long2022MultiCPRAM}. Recently, a number of studies have demonstrated that dense passage retrieval systems based on pre-trained language models (PLMs) are significantly more effective compared to traditional sparse retrieval methods such as BM25~\cite{karpukhin-etal-2020-dense}. To balance efficiency and effectiveness, existing dense passage retrieval methods usually leverage a dual-encoder architecture, where query and passage are encoded into continuous vector representations by PLMs respectively, and then a lightweight score function such as dot product or cosine similarity between two vectors is used to estimate the semantic similarity between the query-passage pair.\footnote{We use the term ``passage'' and ``document'' interchangeably throughout this paper.}

\begin{figure}
    \centering
    \includegraphics[width=0.5\textwidth]{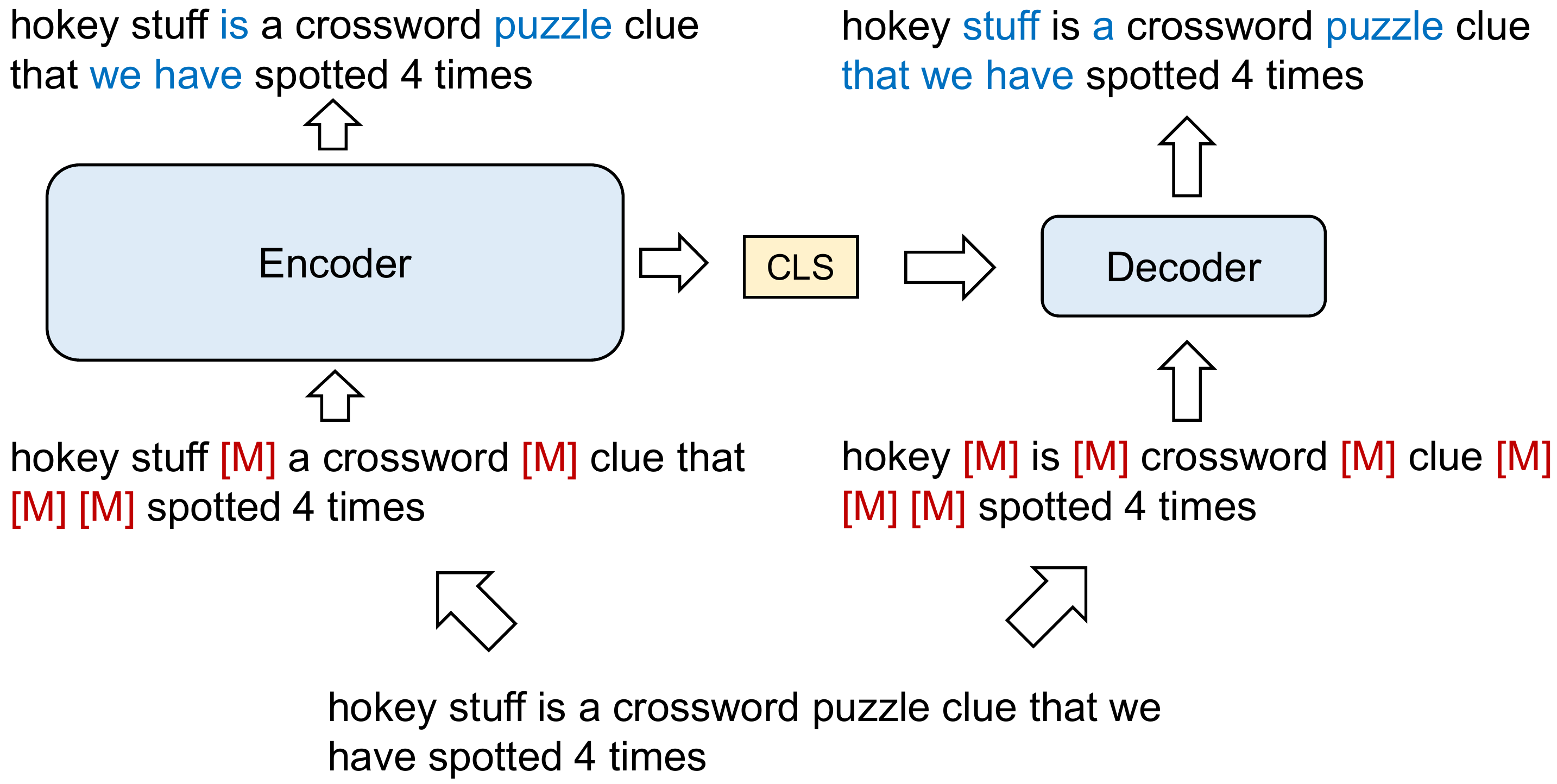}
    \caption{Illustration of the Masked Auto-Encoder (MAE) pre-training framework.}
    \label{fig:mae}
\end{figure}

In the dual-encoder architecture, the text representation capability of the PLMs plays a crucial role as it shall encode all essential information into the low-dimensional dense vector. However, it has been observed that the progress of PLMs in general language understanding benchmarks does not necessarily lead to an improvement in text representation ability~\citep{li-etal-2020-sentence, lu-etal-2021-less, Wang2022SimLMPW} as the widely used masked language modeling (MLM) pre-training objective focuses more on representing individual tokens rather than the entire sentence. As a result, numerous recent studies have explored to enhance the base model's sentence representation ability via incorporating supplementary pre-training tasks or designing new pre-training architectures~\citep{lee-etal-2019-latent, gao-callan-2021-condenser, xiao-etal-2022-retromae}.

Currently, the Masked Auto-Encoder (MAE) is arguably the most effective pre-training framework in retrieval tasks. As illustrated in Figure~\ref{fig:mae}, MAE utilizes the encoder-decoder architecture in which the sentence is randomly masked twice as the input to the encoder and decoder, respectively, and the sentence embedding pooled from the encoder is concatenated with the masked input of the decoder to reconstruct the original input. In this framework, the encoding quality is critical to the success of the reconstruction task of decoder, which in turn, is the key to improving the representation ability of the encoder. To increase the difficulty of the decoder, previous works usually utilized a shallower decoder and a higher decoder mask ratio~\citep{xiao-etal-2022-retromae}. We argue that optimizing the training task of MLM can further enhance the challenge of the decoder, leading to an even greater improvement in the representation ability of the encoder.

\begin{figure}
    \centering
    \includegraphics[width=0.45\textwidth]{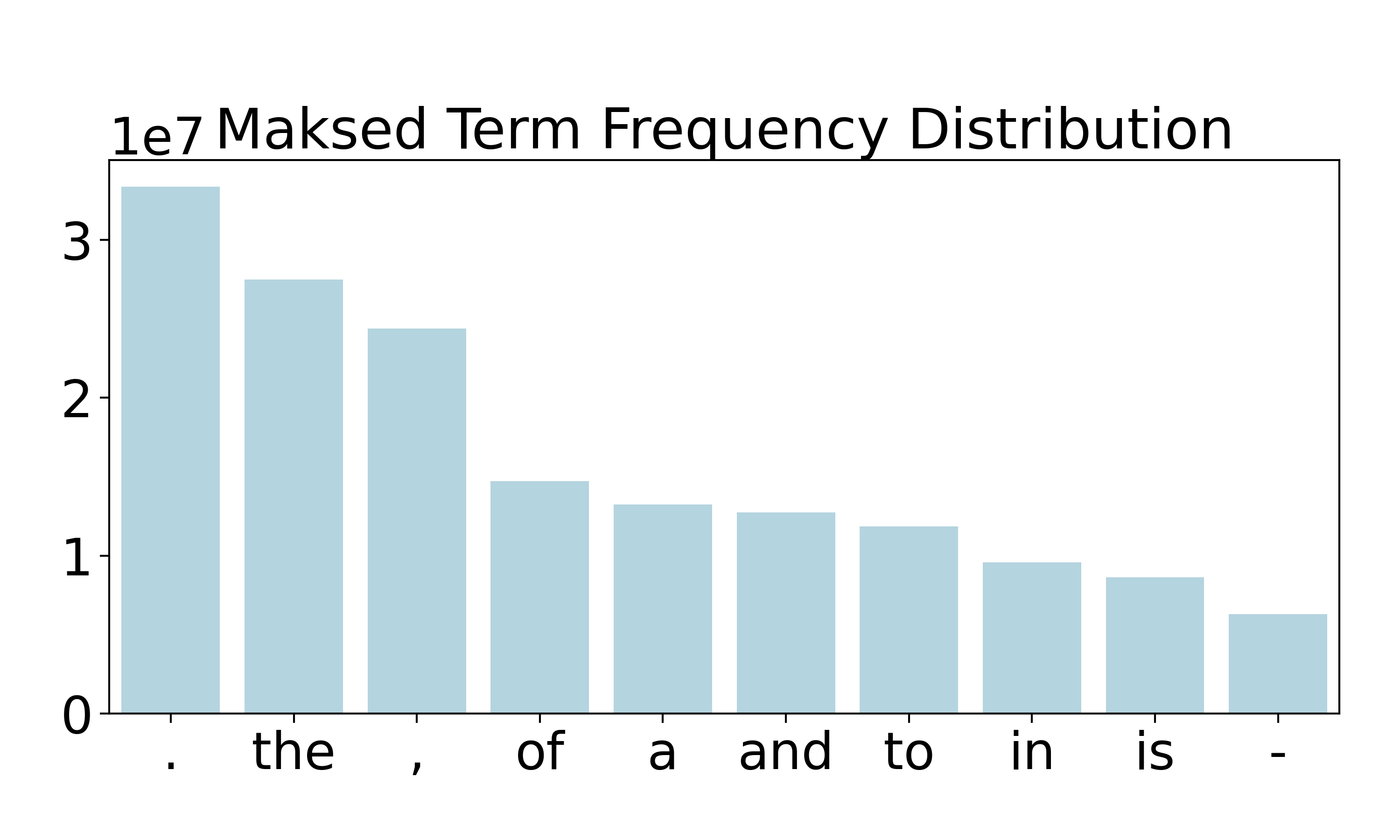}
    \caption{Top 10 masked term frequency distribution using random masking with uniform sampling. Statistics are measured on the MS MARCO passage corpus.}
    \label{fig:top10}
\end{figure}

During the MAE pre-training process, the MLM training objective is used at both the encoder and decoder ends, with masked tokens randomly sampled from a uniform distribution over the input sentence. However, such a random sampling strategy is sub-optimal since it is biased towards sampling uninformative high-frequency tokens (\emph{e.g.}, stop-words and punctuations as shown in Figure~\ref{fig:top10}).\footnote{Uniform sampling from each sentence is statistically equivalent to sampling from the term frequency distribution over the corpus' vocabulary, which is a power-law distribution.}
Previously, many research works have attempted to address this issue by detecting and masking spans~\citep{joshi-etal-2020-spanbert, levine2021pmimasking} or determining which tokens should be selected through the PLMs~\citep{ma2020bprop, long2022retrieval}.
In the information retrieval (IR) community, early lexical retrieval systems~\citep{bm25} also used inverse document frequency for term re-weighting in their score functions.
Some recent works have also suggested that incorporating the term importance information and token-level matching signal with deep contextualized representations~\citep{dai20deepct, khattab20colbert, gao-etal-2021-coil, zhou2022master} is beneficial.

In this paper, we follow similar principles to optimize the MAE pre-training framework for dense retrieval.
Specifically, we aim to increase the proportion of masked tokens that are informative (or meaningful for passage retrieval tasks) during MLM training of the decoder.
To achieve this goal, we introduce pointwise mutual information (PMI) between $n$-grams to estimate the importance of each token, which is subsequently used to regularize the sampling distribution of MLM, increasing the probability of important tokens being masked. More importantly, PMI can be calculated in an unsupervised manner and can model the corpus-wise keywords and co-occurrence features which are valuable for IR tasks.

Compared to previous research which relies on another PLM to determine the importance of tokens~\citep{long2022retrieval} or sample replacements~\citep{Wang2022SimLMPW}, our method is more computationally efficient because it only requires computing statistical information from the corpus.
We apply this masking strategy asymmetrically to the encoder and decoder sides, enforcing the dense bottleneck between them to encode more information about salient co-occurring tokens captured by decoder masking.
We evaluate the effectiveness of our approach on several supervised passage retrieval datasets and zero-shot retrieval benchmarks.

To summarize, our contributions are as follows:

\begin{itemize}
    \item We reveal the problem of uniform sampling in random masking, which biases the learning process of language models toward high-frequency tokens that are uninformative in passage retrieval scenarios.
    \item To mitigate the above issue, we introduce an efficient approach for term importance estimation based on pointwise mutual information.
    We further propose a novel importance-aware masking strategy and integrate this masking strategy with the bottlenecked masked autoencoder framework for better sentence representation learning.
    \item Experiments results demonstrate the effectiveness of our approach. Our model outperforms state-of-the-art dense retrievers in both supervised settings and zero-shot settings.
\end{itemize}

\section{Related Work}

\paragraph{Pre-training for Dense Retrieval}
Dense retrieval usually involves learning a dual-encoder that encodes query and document into dense vectors and takes their inner product (or cosine similarity) as their relevance score.
It's usually trained with the contrastive objective to distinguish positive pairs from negative ones~\citep{karpukhin-etal-2020-dense}.
To reduce the gap between MLM and downstream retrieval tasks, plenty of works have designed various self-supervised contrastive pre-training tasks by automatically constructing positive and negative pairs~\citep{lee-etal-2019-latent, Chang2020Pre-training, izacard2022unsupervised, DBLP:journals/corr/abs-2201-10005}.

Different from the idea of designing pretext contrastive tasks, another line of research has focused on improving the global text representation learning ability in the masked auto-encoding framework.
For example, \citet{lu-etal-2021-less} proposed adding a shallow auto-regressive decoder on top of the encoder for causal language modeling conditioned on the sentence representation.
\citet{gao-callan-2021-condenser} changed the decoder to a bi-directional Transformer head trained with MLM task.
\citet{xiao-etal-2022-retromae} proposed an enhanced-decoding mechanism with token-specific masking design and an aggressive masking ratio.
\citet{Wang2022SimLMPW} introduced ELECTRA-style replaced language modeling to reduce the gap between pre-training and fine-tuning.
In addition to predicting the sentence itself based on its contextualized representation, \citet{wu2022cotmae} proposed to additionally predict its surrounding context.
\citet{zhou2022master} further used docT5query~\citep{docT5query} and GPT-2~\citep{radford2019language} for data augmentation and combined all these tasks with multiple decoders.
It should be noticed that our contributions are orthogonal to theirs since we only use the self-prediction task and the vanilla MAE architecture.

\paragraph{Pointwise Mutual Information}
Using PMI to model word association in NLP was first introduced by ~\citet{church-hanks-1990-word}.
\citet{NIPS2014_feab05aa} found that the skip-gram algorithm is implicitly factorizing the (shifted) word-context PMI matrix for word embedding learning.
More recently, ~\citet{levine2021pmimasking} extended PMI to detect more correlated spans for masking.
\citet{sadeq-etal-2022-informask} identified more informative masking patterns by finding a maximum cut in the PMI matrix.
Different from previous work, we use the PMI to estimate how much a token contributes to the sentence and use the term importance information to regularize the sampling distribution.

\section{Methodology}
In this section, we describe our proposed pre-training method for the dense passage retrieval task. We first give a brief overview of the conventional MAE pre-training architecture. Then we introduce how to extend it to our model transforming the decoder to be more challenging. Finally, we detail the multi-stage fine-tuning and inference designs.

\subsection{Bottlenecked Masked Autoencoders}

In the bottlenecked masked autoencoder (BMAE) architecture, a deep encoder is used to get a compressed vector as sentence embedding, and a shallow decoder is used to recover masked tokens based on the sentence embedding offered by the encoder.
An information bottleneck (\emph{e.g.}, the representation at \verb|[CLS]| position) is imposed as the communication budge between the encoder and decoder.
To make good predictions, the encoder must compress as much information to the bottleneck, whereas the decoder has to attend to the bottleneck representation due to its weakened capacity.

Formally, given a piece of text $\mathbf{x}$, two masking operations are independently applied to the input sequence $\mathbf{x}$ with different masking ratios $p_\text{enc}$ and $p_\text{dec}$. Among the masked tokens, $80\%$ are replaced with a special \texttt{[MASK]} token, $10\%$ are replaced by a random token in the vocabulary, and the remaining tokens are kept unchanged~\citep{devlin-etal-2019-bert},

\begin{equation}
\begin{aligned}
    \hat{\mathbf{x}}_\text{enc} = \text{Mask}(\mathbf{x}, p_\text{enc}), \\
    \hat{\mathbf{x}}_\text{dec} = \text{Mask}(\mathbf{x}, p_\text{dec}).
\end{aligned}
\end{equation}

\paragraph{Deep Encoder}
We use a multi-layer Transformer network~\citep{NIPS2017_3f5ee243} as the deep encoder which can be initialized from PLMs such as BERT~\citep{devlin-etal-2019-bert}.
We feed $\hat{\mathbf{x}}_\text{enc}$ through the deep language model encoder to get its contextualized representations $\mathbf{h}_{\hat{\mathbf{x}}_\text{enc}}$ for masked token reconstruction,

\begin{equation}
    \mathbf{h}_{\hat{\mathbf{x}}_\text{enc}} = \text{Encoder}(\hat{\mathbf{x}}_\text{enc})
\end{equation}

The encoder is trained via the MLM objective where the language modeling head predicts masked tokens based on contextualized representations,

\begin{equation}
    L_\text{enc} = - \sum_{x\in M_\text{enc}} \text{log} P_\text{lm}(x|\mathbf{h}_{\hat{\mathbf{x}}_\text{enc}})
\end{equation}
where $M_\text{enc}$ denotes the set of masked tokens at the encoder side.

\paragraph{Bottleneck Representation}
To ensure consistency with downstream fine-tuning, we use the \verb|[CLS]| token representation $\mathbf{h}_c \in \mathbb{R}^{d_\text{enc}}$ at the last layer of the deep encoder as the text representation.
We use a linear projection head to construct the bottleneck for decoder recovery,
\begin{equation}
    \mathbf{h}_b = \mathbf{W} \mathbf{h}_c + \mathbf{b}
\end{equation}
where $\mathbf{W}\in \mathbb{R}^{d_\text{dec} \times d_\text{enc}} $ and $\mathbf{b} \in \mathbb{R}^{d_\text{dec}}$ are trainable parameters.

\paragraph{Weak Decoder}
The decoder is a shallow neural network consisting of two randomly initialized Transformer layers.
Its objective is to recover the aggressively masked inputs conditioned on the bottleneck representation $\mathbf{h}_b\in \mathbb{R}^{d_\text{dec}}$ provided by the deep encoder,

\begin{equation}
    \mathbf{h}_{\hat{\mathbf{x}}_\text{dec}} = \text{Decoder}(\mathbf{h}_b,\hat{\mathbf{x}}_\text{dec}).
\end{equation}

We use the same language model head for masked language modeling,
\begin{equation}
    L_\text{dec} = - \sum_{x\in M_\text{dec}} \text{log} P_\text{lm}(x|\mathbf{h}_{\hat{\mathbf{x}}_\text{dec}})
\end{equation}
where $M_\text{dec}$ denotes the set of masked tokens at the decoder side.

\subsection{Token Importance Aware Masking}

\newcommand{\textexample}{\hlc[pink!50]{hokey}\ \hlc[pink!85]{stuff} \hlc[pink!0]{is} \hlc[pink!0]{a} \hlc[pink!100]{crossword} \hlc[pink!100]{puzzle} \hlc[pink!100]{clue} \hlc[pink!0]{that} \hlc[pink!0]{we} \hlc[pink!0]{have} \hlc[pink!97]{spotted} \hlc[pink!94]{4} \hlc[pink!50]{times}}

\begin{figure*}
\begin{subfigure}{0.35\textwidth}
    \centering
    \includegraphics[width=\textwidth]{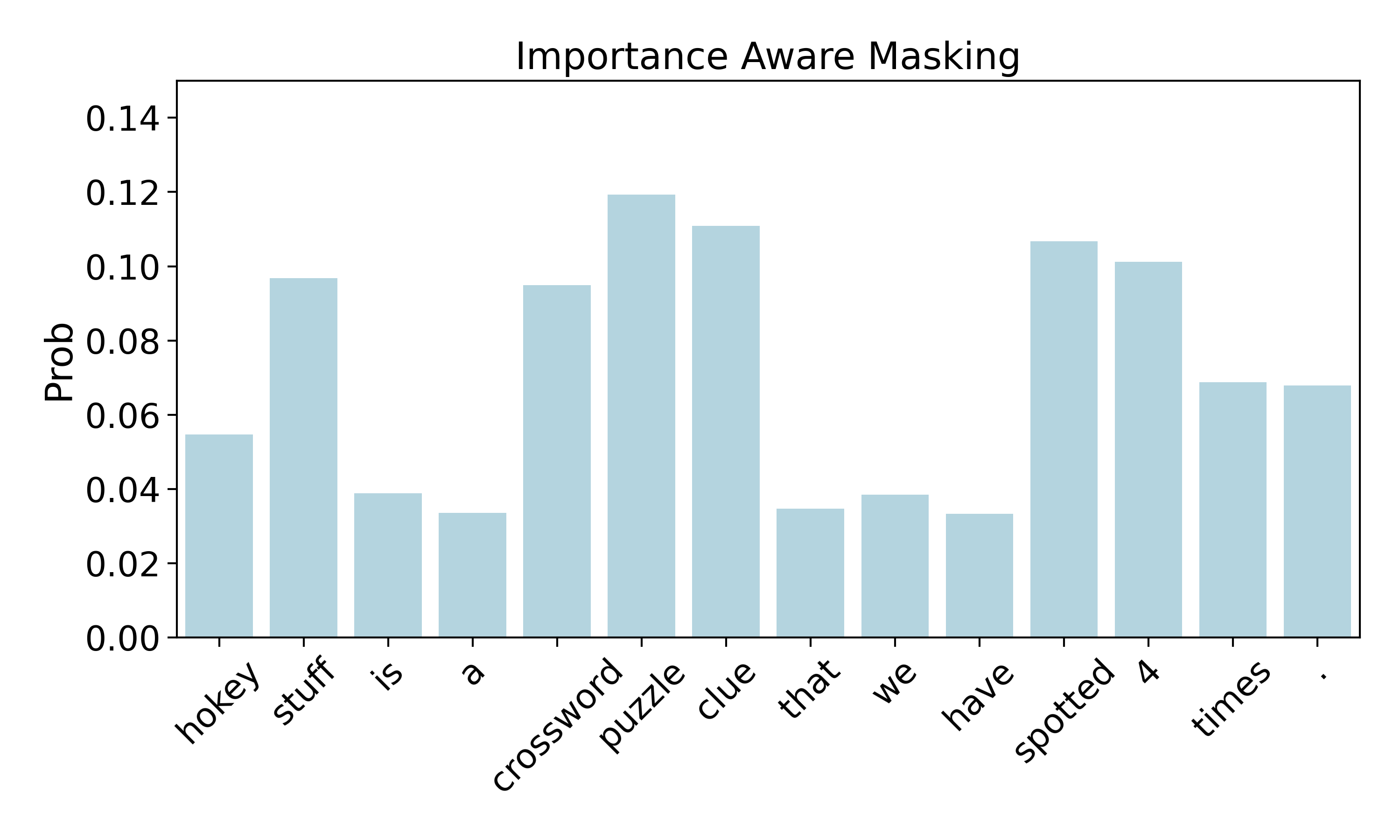} \\
    \includegraphics[width=\textwidth]{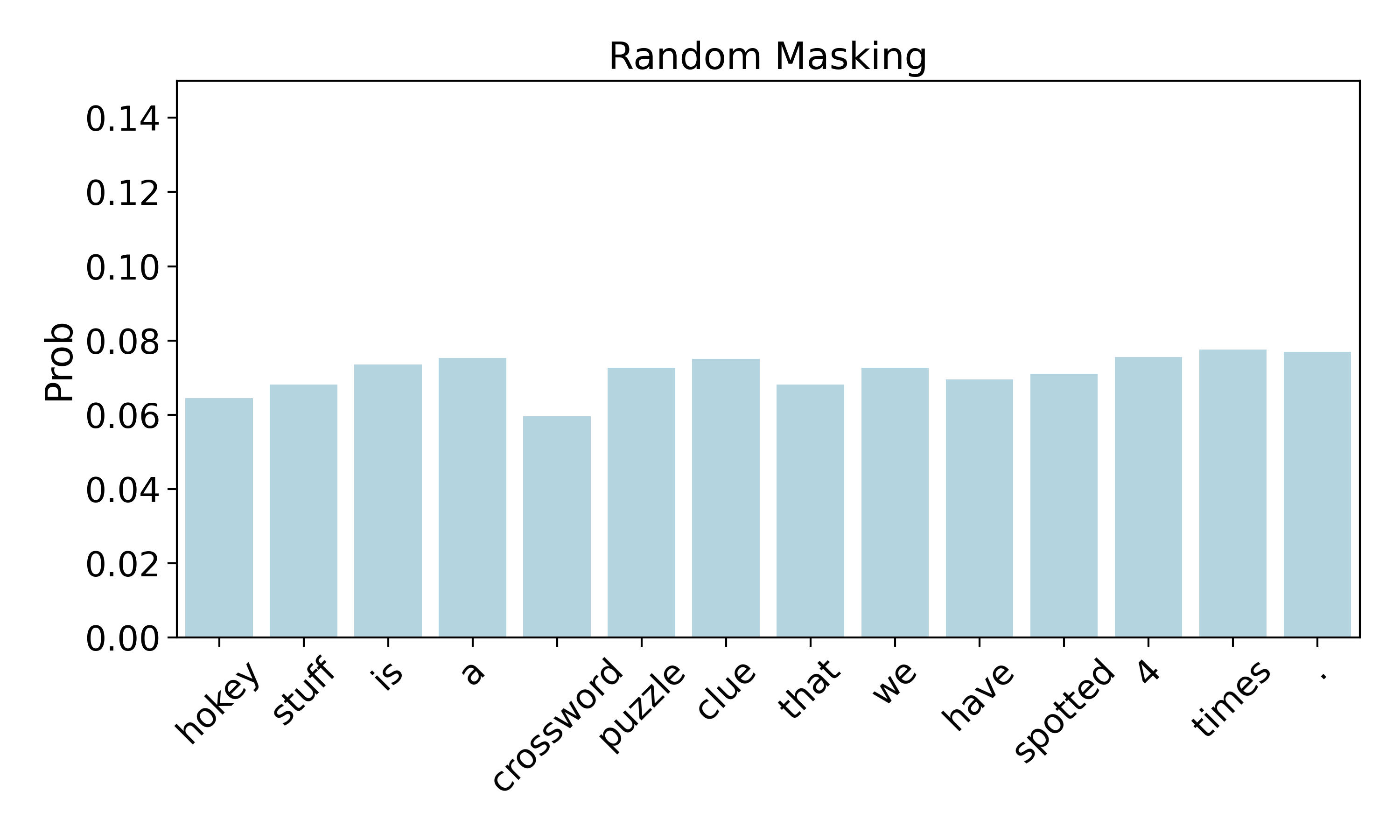}
    \caption{Sampling distribution comparison.}
    \label{fig:sampledist}
\end{subfigure}
\hfill
\begin{subfigure}{0.6\textwidth}
    \centering
    \includegraphics[width=\textwidth]{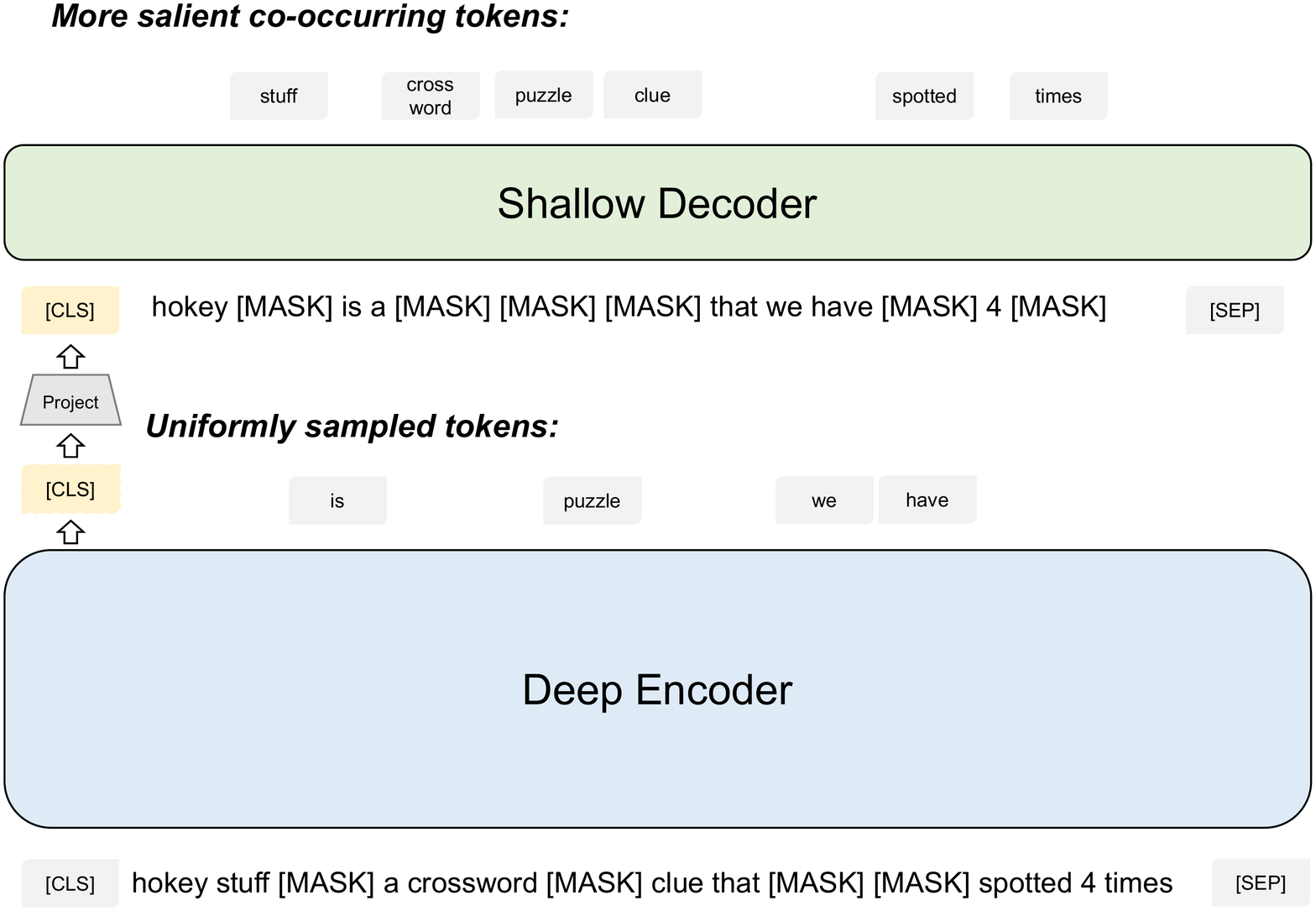}
    \caption{Model architecture.}
    \label{fig:ours}
\end{subfigure}
\caption{Overview of our pre-training approach. We first estimate term importance in an unsupervised manner for a piece of text: \textexample. Then we incorporate the term importance awareness into decoder masking to mask more salient co-occurring tokens.}
\end{figure*}

The random masking strategy treats all tokens equally. Statistical analysis reveals that $40\%$ of the masked tokens produced by the $15\%$ ratio random masking method in BERT pre-training are stop-words or punctuation. However, prior research has shown that the contribution of these tokens to the dense passage retrieval task is limited. Furthermore, these frequently occurring tokens serve to ease the burden on the decoder in the MAE pre-training process. To address the limitation above, we propose a new masking strategy aiming to mask tokens that are more important thus posing a greater challenge for the decoder in the reconstruction task.

\paragraph{Token Importance Estimation} 
Mutual Information is widely used to estimate the correlation of two random variables. Inspired by previous work, we consider linking the importance of tokens to statistical pointwise mutual information. The pointwise mutual information between two words $w_1$ and $w_2$ represents the correlation between them. Mathematically, the pointwise mutual information of the bi-gram combination $w_1w_2$ can be formulated as:
\begin{equation}
    \text{PMI}(w_1, w_2) = log\frac{p(w_1,w_2)}{p(w_1)p(w_2)}
\end{equation}
where the probability of any $n$-gram is defined as the number of its occurrences in the corpus divided by the number of all the n-grams in the corpus.
For each word in a sentence, if the word is more informative (namely more important), then statistically speaking, the PMI between this word and the remaining parts of the sentence should be larger. In practice, calculating the PMI between each word and the remaining parts of the sentence can be computationally expensive and may result in bias. As such, a more suitable approach is to determine the importance of a word by analyzing its PMI with adjacent fragments.

To simplify the calculation without sacrificing performance, we only need to compute the average mutual information (AMI) of fixed window size. In the determined window size, we use the average PMI value of all possible $n$-gram combinations delimited by the token (starting and ending with the token, and with a minimum $n$ value of $2$) as the ultimate reference for token importance estimation. For each word $w_i$ in sentence $\mathcal{S}$, if the window size is set to $L$, the final average mutual information of word $w_i$ can be represented as:
\begin{equation}
\begin{aligned}\label{eq:ami}
\text{AMI}(w_i) & = \frac{1}{L-1}\sum_{j=1}^{L-1}\text{PMI}(w_{i-j},...,w_i) \\
& + \frac{1}{L-1}\sum_{j=1}^{L-1}\text{PMI}(w_i,...,w_{i+j}).
\end{aligned}
\end{equation}
To clarify, we calculate the pointwise mutual information of bigrams when $L=2$. When $L \geq 2$, we will expand to the computation of all $n$-grams where $2\leq n \leq L$. In order to balance computational efficiency and accuracy, we set $L=4$ in this work.

\begin{algorithm}[t]
\caption{Importance Aware Masking.}\label{algo:rom}
\textbf{Input:} $\mathbf{x}=(x_1,\dots,x_n)$, $\mathbf{t}=(t_1,\dots,t_n)$, $p$; \\
\textbf{Output:} $\mathbf{M}=(m_1,\dots,m_n)$;
\begin{algorithmic}[1]
\State $\mathbf{M} \gets []$
\For{$i \textbf{ in } 1,\dots,n$}
    \State{$t \sim \mathcal{N}(\mathbf{t}[i], \sigma)$}
    \State{$\mathbf{t}[i] \gets t$}
\EndFor
\State $k = \lfloor \Call{len}{\mathbf{x}} \times p \rfloor$
\State $\mathbf{I} = \Call{argsort}{\mathbf{t}}$
\For{$i \textbf{ in } 1,\dots,n$}
\If {$i \textbf{ in } \Call{topk}{\mathbf{I},k}$}
    \State {$\Call{append}{\mathbf{M}, 1}$}
\Else
    \State{$\Call{append}{\mathbf{M}, 0}$}
\EndIf
\EndFor
\State \Return $\mathbf{M}$
\end{algorithmic}
\end{algorithm}

\paragraph{Importance Aware Sampling}

Given a sentence $\mathbf{x} = (x_1, x_2, ..., x_n)$ that consists of $n$ tokens equipped with the estimated token importance $\mathbf{t} = (t_1, t_2, ..., t_n)$ where $t_i=\text{AMI}(x_i)$.
Instead of random sampling from a uniform distribution over the sequence of tokens, we propose a novel sampling strategy that takes the term importance into account.
Specifically, we sort tokens based on their perturbed importance and mask tokens with higher importance.
Each token importance is perturbed with a Gaussian noise before being sorted to preserve some randomness during masking.\footnote{We simply set the variance $\sigma=1$.}
Details can be found in Algorithm~\ref{algo:rom}.
The output $\mathbf{M}=(m_1,\dots,m_n)$ is a binary-valued array in which 0 denotes unmasked and 1 for masked.
We apply the proposed masking strategy asymmetrically to the encoder and decoder sides.
The encoder side masking is unchanged while for the decoder, we use the importance aware masking to mask more salient tokens.
Figure~\ref{fig:sampledist} compares the sampling distribution induced by the proposed importance-aware masking algorithm and the uniform sampling distribution of random masking.
More informative words have a higher probability of being masked.

\subsection{Pre-training}

The pre-training is conducted on the unlabeled corpus $\mathcal{C}=\{d_1, d_2, ..., d_N\}$.
The masked auto-encoder is pre-trained with the masked language modeling (MLM) objective from both the encoder and decoder,
\begin{equation}
    L = L_{enc} + L_{dec},
\end{equation}
where each loss is normalized to stabilize training.

\subsection{Fine-tuning}

Given a labeled dataset $\mathcal{D}=\{(q_1,d^+_1), (q_2,d^+_2), ..., (q_n,d^+_n)\}$ consisting of query and supporting document pairs, we adopt the fine-tuning pipeline from prior work~\cite{Wang2022SimLMPW}. The fine-tuning consists of three stages: the first retriever is trained with the official BM25 hard negatives; the second retriever is trained with the hard negatives mined using the first retriever; the final stage retriever is trained with the hard negatives mined via the second retriever and knowledge distillation from a cross-encoder reranker.
At each of the three training stages, retrievers are initialized with the same pre-trained checkpoint.

\paragraph{Stage 1: BM25 Negatives.}
For each query $q$,  we sample hard negatives $\mathbb{D}^-_\text{BM25}$ from top $K_1$ ranked documents excluding the positive ones returned by BM25.
In-batch negatives are also included to improve training efficiency.
The retriever is trained as a dual encoder with a contrastive loss,
\begin{equation} \label{equ:infonce}
    L_{\text{cont}} = -\log \frac{\text{exp}(s(q, d^+))}{\sum_{d\in \{d^+\}\cup \mathbb{D}^-_\text{BM25}}\text{exp}(s(q, d))}
\end{equation}
where $s(q, d)$ measures the similarity score between query $q$ and document $d$.
In this paper, we use the dot product between query and document representations as their similarity score $s_\text{de}(q,d) = \mathbf{q} \cdot \mathbf{d}$, where $\mathbf{q} = \text{DE}(q)$ and $\mathbf{d} = \text{DE}(d)$.

\paragraph{Stage 2: Hard Negatives.}
In the second stage, we substitute the BM25 lexical retriever with the dense retriever trained in the first stage for mining more challenging hard negatives~\cite{xiong2021approximate}.
Top $K_2$ documents excluding positives are collected to construct $\mathbb{D}^-_\text{hard}$.
The retriever is trained with the same objective as Equation \ref{equ:infonce} by substituting $\mathbb{D}^-_\text{BM25}$ with $\mathbb{D}^-_\text{hard}$.

\paragraph{Stage 3: Reranker-Distilled.}
Hard negatives mined by retrievers suffer from the problem of false negatives due to the incomplete data annotation, which can be relieved by pseudo-labeling from a more powerful yet costly cross-encoder reranker~\citep{qu-etal-2021-rocketqa, ren-etal-2021-rocketqav2}.
The cross-encoder takes the concatenation of query and document as input and models their relevance through multi-layer token-level interactions, making it a better architecture for fine-grained relevance estimation.
It is trained via the list-wise contrastive loss similar to the dual-encoder retriever except that the score function becomes a parametric form $s_\text{ce}(q,d)=\text{CE}(\text{concat}(q,d))$ and in-batch negatives are not involved in loss computation.
The hard negatives $\mathbb{D}^-_\text{hard$^2$}$ are mined from the top $K_3$ documents returned by the previous stage retriever.

We use the KL (Kullback-Leibler) divergence between the relevance distribution estimated by the dual-encoder retriever and the cross-encoder reranker for knowledge distillation:
\begin{equation}
\begin{aligned}
    L_\text{kl} = \text{KL}(P_\text{ce}(D|q)||P_\text{de}(D|q)), \\
    = \sum_{d\in D} p_\text{ce}(d|q) \log \frac{p_\text{ce}(d|q)}{p_\text{de}(d|q)}.
\end{aligned}
\end{equation}
where $p(d|q) \propto \text{exp}(s(q,d))$, $D$ consists of the positive document $d^+$ and a set of negative documents sampled from $\mathbb{D}^-_\text{hard$^2$}$.
The original contrastive loss is also added following~\citet{Wang2022SimLMPW}.
The final objective during knowledge distillation stage is $L = L_\text{kl} + \alpha L_\text{cont}$.

\subsection{Inference}

The inference for a dense retriever is conducted in two stages: index and retrieval.
The document encoder first encodes each document in the corpus into a fix-dimensional dense vector over which a dense index is built.
Given a query $q$, query encoder is employed to get the query representation, and retrieval is done by performing a maximum inner product search over the dense index to find relevant documents from the corpus.

\section{Experiments}
\begin{table*}[!ht]
\setlength{\tabcolsep}{3pt}
\centering
\resizebox{\textwidth}{!}{
\begin{tabular}{lccccccc}
\toprule
\multicolumn{1}{c}{\multirow{2}{*}{\textbf{Method}}} & \multirow{2}{*}{\textbf{\makecell{Reranker \\ distilled}}} &  \multirow{2}{*}{\textbf{\makecell{Multi \\ Vec}}} & \multicolumn{3}{c}{\textbf{MS MARCO dev}} & \multicolumn{1}{c}{\textbf{TREC DL 19}} & \multicolumn{1}{c}{\textbf{TREC DL 20}} \\
\cmidrule(lr){4-6} \cmidrule(l){7-7} \cmidrule(l){8-8}
\multicolumn{1}{c}{} & \multicolumn{1}{c}{}  & & MRR@10 & Recall@50 & Recall@1k & nDCG@10  & nDCG@10 \\
\hline\hline
BM25~\citep{bm25}  &  & & 18.5 &  58.5   &  85.7   &   51.2  &  47.7   \\
DeepCT~\citep{dai20deepct} &  &  &  24.3  & 69.0 &  91.0   &  57.2   &   -   \\
docT5query~\citep{docT5query} &  &  &  27.7  &  75.6  & 94.7  &  64.2  &  -  \\
\hline
ANCE~\citep{xiong2021approximate} &  &  &  33.0 & -  &  95.9  &   64.5  & 64.6 \\
SEED~\citep{lu-etal-2021-less} &  &  & 33.9 &  -  &  96.1  &   -  & - \\
TAS-B~\citep{hofstatter21tasb} & \cmark & &  34.0 &  -  &  97.5  &  71.2  & 69.3  \\
ColBERT~\citep{khattab20colbert} & & \cmark &   36.0  &   82.9  &  96.8   &  -  & - \\
COIL~\citep{gao-etal-2021-coil} & & \cmark & 35.5 & - & 96.3 & 70.4 & - \\
Condenser~\citep{gao-callan-2021-condenser} & &  &  36.6  &   -   &  97.4   &  69.8   &  -  \\
RocketQA~\citep{qu-etal-2021-rocketqa} & \cmark & &  37.0   &   85.5  &   97.9   &  -  & -  \\
PAIR~\citep{ren-etal-2021-pair} & \cmark & & 37.9  &  86.4  &  98.2   &  -  & -  \\
coCondenser~\citep{gao-callan-2022-unsupervised} & & &  38.2   &  86.5   &  98.4  &  71.7  &  68.4  \\
RocketQAv2~\citep{ren-etal-2021-rocketqav2} &  \cmark & &  38.8  &  86.2    &   98.1   &  - &  - \\
AR2~\citep{zhang2022adversarial} &  \cmark  & &   39.5  &  87.8  &  98.6   &  - &  - \\
ColBERTv2~\citep{santhanam-etal-2022-colbertv2} &  \cmark  & \cmark & 39.7 &  86.8 & 98.4 & - & - \\
UnifieR$_\text{dense}$~\citep{shen21unifier} & & & 38.8 & - & 97.6 & 71.1 & - \\
\hline
RetroMAE~\citep{xiao-etal-2022-retromae} & \cmark & & \underline{41.6} & 88.5 & \underline{98.8} & 68.1 & 70.6 \\
SimLM~\citep{Wang2022SimLMPW} & \cmark & & 41.1 & 87.8 & 98.7 & 71.2 & 69.7 \\
CoT-MAE~\citep{wu2022cotmae} & & & 39.4 & 87.0 & 98.7 & - & 70.4 \\
MASTER~\citep{zhou2022master} & \cmark & & 41.5 & \textbf{88.6} & \underline{98.8} & \underline{72.7} & \textbf{71.7} \\
\hline
\ours & \cmark & & \textbf{41.7} & \textbf{88.6} & \textbf{98.9} & \textbf{73.0} & \underline{71.1} \\
\hline
\bottomrule
\end{tabular}
}
\caption{
Main results on MS MARCO passage ranking and TREC Deep Learning tracks.
}
\label{tab:ir_results}
\end{table*}

\begin{table*}[!ht]
\centering
\setlength{\tabcolsep}{3pt}
\resizebox{\textwidth}{!}{
\begin{tabular}{lccccccc}
\toprule
\multicolumn{1}{c}{\multirow{2}{*}{\textbf{Method}}} & \multirow{2}{*}{\textbf{\makecell{Pretraining \\ Setting}}} & \multicolumn{2}{c}{\textbf{BM25 Negatives}}                  & \multicolumn{2}{c}{\textbf{Hard Negatives}} & \multicolumn{2}{c}{\textbf{Reranker-Distilled}}                                   \\ \cmidrule(l){3-4} \cmidrule(l){5-6} \cmidrule(l){7-8}
\multicolumn{1}{c}{} & & MRR@10     & Recall@1k & MRR@10     & Recall@1k       & MRR@10     & Recall@1k  \\ \midrule
coCondenser~\citep{gao-callan-2022-unsupervised} & unsupervised & 35.7 & 97.8 & 38.2 & 98.4 & 40.2 & 98.3 \\
RetroMAE~\citep{xiao-etal-2022-retromae} & unsupervised & 37.7 & 98.5 & 39.3 & 98.5 & \underline{41.6} & \underline{98.8} \\
SimLM~\citep{Wang2022SimLMPW} & unsupervised & 38.0 & 98.3 & 39.1 & 98.6 & 41.1 & 98.7 \\
CoT-MAE~\citep{wu2022cotmae} & unsupervised & 36.8 & 98.3 & 39.2 & \underline{98.7} & 40.2 & 98.3 \\
MASTER~\citep{zhou2022master} & semi-supervised & \underline{38.3} & \textbf{98.8} & \textbf{40.4} & \textbf{98.8} & 41.5 & \underline{98.8} \\
\hline
\ours & unsupervised & \textbf{38.8} & \underline{98.6} & \underline{40.2} & \underline{98.7} & \textbf{41.7} & \textbf{98.9} \\
\bottomrule
\end{tabular}
}
\caption{
Performance at different stages of the fine-tuning pipeline on MS MARCO dev set.
Other model results are borrowed from the corresponding papers.
We rerun RetroMAE first stage results using their code since it is not reported in their paper.
} \label{tab:multi_stage_res}
\end{table*}

\subsection{Setup}

The pre-training consists of two stages: generic pre-training followed by unsupervised corpus-aware pre-training.
\footnote{
Due to resource limitation, we omit the pre-training on general corpus by initializing from retromae checkpoint pre-trained on the combination of Wikipedia and BookCorpus and continually pre-train it on the MS MARCO passage corpus consisting of 8.8M passages.
}
We set encoder mask ratio $p_\text{enc}=0.3$ and decoder mask ratio $p_\text{dec}=0.5$. 
After pre-training, the decoder is discarded and only the encoder is used for retriever initialization.
During fine-tuning, we use a siamese architecture that shares the parameters of query encoder and document encoder.
We set hard negative mining depths $K_1=1000$, $K_2=K_3=200$, and contrastive loss ratio $\alpha=0.2$ following~\citet{Wang2022SimLMPW}.
Other hyperparameters are adopted from prior work and detailed in Appendix~\ref{sec:appendix}.
During inference, we use the \verb|faiss| library~\citep{faiss} to build the index and perform the exact search.

We evaluate our approach in two settings: supervised in-domain passage retrieval and zero-shot out-of-domain retrieval.
For supervised in-domain evaluation, we choose MS MARCO passage retrieval task~\citep{DBLP:journals/corr/NguyenRSGTMD16}, TREC 2019 and 2020 Deep Learning Tracks~\citep{DBLP:journals/corr/abs-2003-07820, DBLP:conf/trec/CraswellMMYC20}.
For zero-shot evaluation, we report results on 14 open available datasets of BEIR benchmark~\citep{nandan21beir}.

We mainly compare our approach to other pre-training approaches tailored for dense retrieval, including Condenser~\citep{gao-callan-2021-condenser}, coCondenser~\citep{gao-callan-2022-unsupervised}, RetroMAE~\citep{xiao-etal-2022-retromae}, SimLM~\citep{Wang2022SimLMPW}, CoT-MAE~\citep{wu2022cotmae}, MASTER~\citep{zhou2022master}. We also list the performance of other retrieval systems for reference.

\subsection{Main Results}

\begin{table*}[t]
\centering
\setlength{\tabcolsep}{3pt}
\resizebox{\textwidth}{!}{
\begin{tabular}{lccccccccc}
\toprule
\textbf{Method} & BM25 & DPR & ColBERT & Contriever & Condenser & RetroMAE & SimLM & MASTER & \ours \\
 \midrule
TREC-COVID & 65.6 & 33.2 & 67.7 & 59.6 & 75.0 & \underline{77.2} & 63.7 & 62.0 & \textbf{79.0} \\
NFCorpus & 32.5 & 18.9 & 30.5 & 32.8 & 27.7 & 30.8 & 32.3 & \underline{33.0} & \textbf{33.4} \\
NQ & 32.9 & 47.4 & \underline{52.4} & 49.8 & 48.6 & 51.8 & 47.7 & 51.6 & \textbf{55.9} \\
HotpotQA & 60.3 & 39.1 & 59.3 & \underline{63.8} & 53.8 & 63.5 & 58.1 & 58.9 & \textbf{65.5} \\
FiQA-2018 & 23.6 & 11.2 & 31.7 & \underline{32.9} & 25.9 & 31.6 & 29.2 & 32.8 & \textbf{33.6} \\
ArguAna & 31.5 & 17.5 & 23.3 & \underline{44.6} & 29.8 & 43.3 & 42.1 & 39.5 & \textbf{49.1} \\
Touche-2020 & \textbf{36.7} & 13.1 & 20.2 & 23.0 & 24.8 & 23.7 & 29.2 & \underline{32.0} & 21.1 \\
CQADupStack & 29.9 & 15.3 & \underline{35.0} & 34.5 & 34.7 & 31.7 & 33.2 & 32.7 & \textbf{36.1} \\
Quora & 78.9 & 24.8 & 85.4 & \textbf{86.5} & 85.3 & 84.7 & 77.3 & 79.1 & \underline{86.0} \\
DBPedia & 31.3 & 26.3 & 39.2 & \underline{41.3} & 33.9 & 39.0 & 34.5 & 39.9 & \textbf{42.0} \\
SCIDOCS & 15.8 & 7.7 & 14.5 & \underline{16.5} & 13.3 & 15.0 & 14.5 & 14.1 & \textbf{17.0} \\
FEVER & 75.3 & 56.2 & \underline{77.1} & 75.8 & 69.1 & \textbf{77.4} & 65.7 & 69.2 & 76.1 \\
Climate-FEVER & 21.3 & 14.8 & 18.4 & \textbf{23.7} & 21.1 & 23.2 & 16.3 & 21.5 & \underline{23.6} \\
SciFact & 66.5 & 31.8 & \underline{67.1} & \textbf{67.7} & 59.3 & 65.3 & 58.8 & 63.7 & \underline{67.1} \\
\midrule
Best on & 1 & 0 & 0 & \underline{3} & 0 & 1 & 0 & 0 & \textbf{9} \\
Average & 43.0 & 25.5 & 44.4 & 46.6 & 43.0 & \underline{47.0} & 43.0 & 45.0 & \textbf{49.0} \\
\bottomrule
\end{tabular}
}
\caption{Zero-shot transfer performance (nDCG@10) on BEIR benchmark. The best score on a given dataset is marked in \textbf{bold}, and the second best is \underline{underlined}.
Baseline results are taken from the corresponding papers.}

\label{tab:exp_beir_details}
\end{table*}

\paragraph{In-domain Evaluation.}

Results on MS MARCO passage retrieval dataset and TREC DL benchmarks are shown in Table~\ref{tab:ir_results}.
In line with previous study, we find that corpus-aware pre-training brings substantial improvements over sophisticated fine-tuning techniques, such as multi-vector encoding and data augmentations.
Among recently proposed dense retrieval pre-training approaches, our model achieves the best results on MS MARCO and TREC DL 2019, even outperforming MASTER, a semi-supervised pre-training approach that uses DocT5Query~\citep{docT5query, DBLP:journals/corr/abs-1904-08375} generated queries for additional data augmentation.

Table~\ref{tab:multi_stage_res} illustrates the retrieval performance at different fine-tuning stages.
When fine-tuning with only BM25 negatives, our model achieves the best MRR@10 over all pre-training baselines, outperforming RetroMAE by 1.1 points.
When training with self-mined hard negatives, our model outperforms previous unsupervised baselines by a large margin and is competitive to MASTER which pre-trained longer using more supervision signals.
When using knowledge distillation from a cross-encoder reranker, our model achieves the best overall results.
We also observe that the improvement diminishes with the incorporation of knowledge distillation due to more accurate labeling.
However, running through the cross-encoder reranker for each training instance is costly.
Our model is more robust to label noise presented in the training data given its strong performance in the first two stages.

In terms of retrieval efficiency, our model uses a dual-encoder architecture which brings no extra inference overhead.

\paragraph{Zero-shot Evaluation.}

Next, we evaluate the out-of-domain robustness of our pre-training approach on the BEIR benchmark.
We apply the same pre-training procedure on the unlabeled corpus and use the in-domain supervised training data from the BEIR benchmark for fine-tuning.
Table~\ref{tab:exp_beir_details} shows the zero-shot retrieval performance of the retriever on 14 out-of-domain datasets when finetuned with only in-domain data from MS MARCO.
We observe that the contrastively pre-trained Contriever is a strong baseline, which outperforms SimLM and MASTER, but underperforms RetroMAE by a small margin.
Our model shows the best overall performance, outperforming MASTER by 4 points of nDCG@10 and RetroMAE by 2 points, performing best on 9 out of 14 datasets.
This verifies the strong generalization ability of our pre-training approach.

\subsection{Ablations and Analysis}
In this section, we ablate and analyze several design choices adopted in model training.
By default, we report models' downstream retrieval performance when fine-tuned using only BM25 hard negatives.\footnote{
MRR@10 on MS MARCO dev (MS dev) and nDCG@10 on TREC DL tracks are used for ablation evaluation.
}



\paragraph{Projection Head.}
Adding a projection head to transform the representation space is commonly used in contrastive representation learning~\citep{pmlr-v119-chen20j}.
We investigate whether this similar approach can contribute to the representation learning in the MAE framework.
Table~\ref{tab:proj} compares the downstream fine-tuning retrieval performance with and without the linear projection header between the encoder and decoder during pre-training.
We observe that adding a linear projection head consistently improves the downstream retrieval performance.
In the MAE framework, the projection head acts as a task adapter that can extract task-specific information from hidden representations.
Therefore, the representation bottleneck before the projection head can focus more on learning task-agnostic information which can generalize better to downstream tasks.

\begin{table}[!ht]
    \centering
    \begin{tabular}{rccc}
    \hline
    Setting & MS dev & DL 19 & DL 20 \\
    \hline\hline
    w/o projection & 38.4 & 66.3 & 65.9 \\
    w/ projection & 38.8 & 67.0 & 67.9 \\
    \hline
    \end{tabular}
    \caption{Ablation of the projection head.}
    \label{tab:proj}
\end{table}

\paragraph{Pre-training Corpus.}
Pre-training language models on the general corpus or continuous pre-training on domain-specific unlabeled corpus are beneficial for downstream domain-specific tasks~\citep{gururangan-etal-2020-dont, xiao-etal-2022-retromae, Wang2022SimLMPW}.
An unanswered question is whether pre-training on the general corpus is still beneficial in the presence of domain-specific corpus.
To validate this, we ablate the MAE-style pre-training on the general corpus by changing the initialization checkpoint before pre-training and fine-tuning it on the MS MARCO corpus.
Table~\ref{tab:corpus} illustrates the downstream fine-tuning retrieval performance when initialized from different pre-trained checkpoints.
We find that adding a warmup pre-training phase on the general corpus before pre-training on domain-specific corpus is still beneficial to downstream tasks.

\begin{table}[!ht]
    \centering
    \begin{tabular}{cccc}
    \hline
    Corpus & MS dev & DL 19 & DL 20 \\
    \hline\hline
    MS & 37.9 & 64.4 & 67.6 \\
    General $\rightarrow$ MS & 38.8 & 67.0 & 67.9 \\
    \hline
    \end{tabular}
    \caption{Ablation of pre-training corpus.}
    \label{tab:corpus}
\end{table}

\begin{figure}
    \centering
    \begin{tikzpicture}
    \begin{axis}[
    xmode=log,
    log basis x=2,
    ymin = 30, ymax = 40,
    xlabel={$N$},
    ylabel={\small MRR@10},
    minor tick num = 1,
    width = 0.4\textwidth,
]
\addplot [blue, mark=square] coordinates {
    (2,35.3)
    (4,36.9)
    (8,37.4)
    (16,38.2)
    (32,38.4)
    (64,38.4)
};
\label{plot_one}
\end{axis}

\begin{axis}[
    xmode=log,
    log basis x=2,
    ymin = 62, ymax = 72,
    xlabel={$N$},
    ylabel={\small nDCG@10},
    axis y line*=right,
    ylabel near ticks,
    minor tick num = 1,
    width = 0.4\textwidth,
    legend cell align = {left},
    legend pos = south east,
]
\addlegendimage{/pgfplots/refstyle=plot_one}\addlegendentry{MS MARCO}
\addplot [green, mark=o] coordinates {
    (2,66.0)
    (4,66.9)
    (8,66.3)
    (16,67.1)
    (32,66.4)
    (64,67.5)
};
\addlegendentry{TREC DL}
\end{axis}
    \end{tikzpicture}
    \caption{Analysis on the number of hard negatives used in fine-tuning on MS MARCO dev set and TREC DL tracks (averaged).}
    \label{fig:hn}
\end{figure}
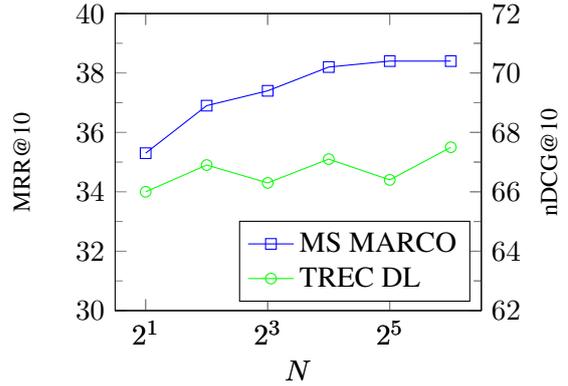

\paragraph{Scaling Hard Negatives.}
More and harder negatives help in contrastive learning by providing a better gradient estimation but suffer from the problem of false negatives~\citep{xiong2021approximate, qu-etal-2021-rocketqa}.
We empirically study the effect of the number of hard negatives used in retriever training.
In a training configuration of batch size $B$ and group size $N$, the total number of negatives for a given query is $B\cdot N - 1$, where $N-1$ are hard negatives and the remaining $(B-1)\cdot N$ are in-batch random negatives.
We keep the total number of negatives $B \cdot N - 1$ fixed and scale the proportion of hard negatives by changing $N$ and $B$ accordingly.
Results are shown in Figure~\ref{fig:hn}.
We find that even without denoising false negatives, keep increasing the number of hard negatives during fine-tuning generally leads to better performance.

\paragraph{Masking Latency.}
We list the theoretical complexity and empirical time latency of different masking operations adopted by various models in Table~\ref{tab:mask-latency}.
Empirical time consumptions are measured at max sequence length $n=150$ and batch size of 128 on one V100 GPU card.
One advantage of our masking strategy is that it is more efficient compared to previous approaches.
In contrast, RetroMAE~\citep{xiao-etal-2022-retromae} utilizes token-specific masks which involve sampling a mask for each token; SimLM~\citep{Wang2022SimLMPW} needs two extra forward computations of an ELECTRA generator for sampling masked token replacements.
This poses no problem for short sequences but is potentially problematic for longer sequences.

\begin{table}[ht]
    \centering
    \begin{tabular}{lcrr}
    \hline
    Model & Complexity & CPU & GPU \\
    \hline
    MAE & $O(n)$ & 0.4ms & n.a \\
    RetroMAE & $O(n^2)$ & 11.6ms & n.a \\
    SimLM & $O(n^2)$ & 0.4ms & 2.8ms \\
    \ours & $O(n\cdot logn)$ & 1.1ms & n.a \\
    \hline
    \end{tabular}
    \caption{Time complexity comparison of different masking strategies adopted in various pre-training approaches. $n$ is the max sequence length.}
    \label{tab:mask-latency}
\end{table}




\section{Conclusion}
In this paper, we propose a novel token importance-aware masking strategy to sample more salient co-occurring tokens based on pointwise mutual information, which can be implemented efficiently in an unsupervised manner.
This masking strategy is applied asymmetrically to the encoder and decoder side for better bottleneck representation learning.
Experiments on both in-domain and zero-shot retrieval benchmarks demonstrate the effectiveness of our method, outperforming recently proposed dense retrieval pre-training approaches.


\bibliography{anthology,custom}
\bibliographystyle{acl_natbib}

\appendix

\section{Implementation Details}
\label{sec:appendix}

For the pre-training stage, we use a learning rate of $3\times10^{-4}$ with linear scheduling, total batch size of 2048, 0.1 warmup ratio, and AdamW optimizer with 0.01 weight decay.
We run the pre-training for 20 epochs and take the checkpoint at $80k$ steps for retriever initialization.
For supervised fine-tuning, we adopt the three-stage fine-tuning procedure.
The first two-stage retrievers are trained for 3 epochs with a peak learning rate of $2\times 10^{-5}$ with linear scheduling, global batch size of 64, and train group size of 32 in which one is positive and the others are negatives.
We swap the batch size and group size in the third stage.
Reranker is initialized from ELECTRA-base following~\cite{Wang2022SimLMPW}.
We run all experiments on 8 NVIDIA Tesla V100 GPUs with 32GB memory.
We fix the random seed as 42 for reproducibility.

\end{document}